\documentclass[journal=jacsat,manuscript=article]{achemso}
\usepackage{amsmath,amsfonts}
\usepackage{booktabs}
\usepackage{color}
\usepackage{algorithm}
\usepackage{algpseudocode}
\usepackage{makecell}
\usepackage{url}
\usepackage{chemformula} 
\usepackage[T1]{fontenc} 

\author{Yun Li}
\affiliation{Department of Chemical and Biological Engineering, University of British Columbia, Vancouver, BC, Canada}
\author{Yixiu Wang}
\affiliation[UBC]{Department of Chemical and Biological Engineering, University of British Columbia, Vancouver, BC, Canada}
\author{Yifu Chen}
\affiliation[UW]{Department of Chemical and Biological Engineering, University of Wisconsin-Madison, Madison, WI, USA}
\author{Kaixun Hua}
\affiliation[UBC]{Department of Chemical and Biological Engineering, University of British Columbia, Vancouver, BC, Canada}
\author{Jiayang Ren}
\affiliation[UBC]{Department of Chemical and Biological Engineering, University of British Columbia, Vancouver, BC, Canada}
\author{Ghazaleh Mozafari}
\affiliation[UBC]{Department of Chemical and Biological Engineering, University of British Columbia, Vancouver, BC, Canada}
\author{Qiugang Lu}
\affiliation[TTU]{Department of Chemical Engineering, Texas Tech University, Lubbock, TX, USA}
\author{Yankai Cao}
\affiliation[UBC]{Department of Chemical and Biological Engineering, University of British Columbia, Vancouver, BC, Canada}
\email{yankai.cao@mail.ubc.ca}
\affiliation[UBC]{Department of Chemical and Biological Engineering, University of British Columbia, Vancouver, BC, Canada}

\title[An \textsf{achemso} demo]
  {Deep Learning-based Predictive Control of Battery Management for Frequency Regulation}

\abbreviations{IR,NMR,UV}
\keywords{American Chemical Society, \LaTeX}

\begin{document}

\begin{abstract}
  This paper proposes a deep learning-based optimal battery management scheme for frequency regulation (FR) by integrating model predictive control (MPC), supervised learning (SL), reinforcement learning (RL), and high-fidelity battery models. By taking advantage of deep neural networks (DNNs), the derived DNN-approximated policy is computationally efficient in online implementation. The design procedure of the proposed scheme consists of two sequential processes: (1) the SL process, in which we first run a simulation with an MPC embedding a low-fidelity battery model to generate a training data set, and then, based on the generated data set, we optimize a DNN-approximated policy using SL algorithms; and (2) the RL process, in which we utilize RL algorithms to improve the performance of the DNN-approximated policy by balancing short-term economic incentives and long-term battery degradation. The SL process speeds up the subsequent RL process by providing a good initialization. By utilizing RL algorithms, one prominent property of the proposed scheme is that it can learn from the data generated by simulating the FR policy on the high-fidelity battery simulator to adjust the DNN-approximated policy, which is originally based on low-fidelity battery model. A case study using real-world data of FR signals and prices is performed. Simulation results show that, compared to conventional MPC schemes, the proposed deep learning-based scheme can effectively achieve higher economic benefits of FR participation while maintaining lower online computational cost.
\end{abstract}

\section{Introduction}
In recent years, an increased number of battery systems are injected into the power grid to provide flexibility to energy systems and mitigate the influences of the increasing penetration of intermittent renewable energy resources. One important role of the battery system in the power grid is to provide frequency regulation (FR) for independent system operators (ISOs) \cite{kim2016data,oudalov2006value}. FR aims at balancing power generation and load demand to maintain the frequency of the power grid around its nominal value. By participating in FR market, the battery system receives compensation from ISOs according to given FR prices. In the process of FR participation, the battery system will also suffer from capacity fade due to aggressively charging/discharging batteries induced by FR signals. For the battery system, determining the optimal FR strategy to maximize the profit of FR participation is a nontrivial task due to the following considerations. On the one hand, this decision-making problem is multi-scale in nature; the battery system needs to consider short-term fluctuations of FR signals and prices while mitigating long-term battery aging. On the other hand, uncertainties, such as un-modeled battery dynamics and uncertain signals in FR, can bring additional challenges for designing optimal FR strategies.

Optimal strategies for battery participation in FR markets have received much attention and are still being investigated \cite{he2015optimal,xu2018optimal,sadeghi2021optimal,kumar2018stochastic,kumar2019benchmarking}. Especially, as a powerful control framework for optimizing control performance while guaranteeing system constraints, model predictive control (MPC) is utilized in designing optimal battery management strategies for FR. \citeauthor{kumar2018stochastic}\cite{kumar2018stochastic} proposed a stochastic MPC scheme combining uncertainty qualification (UQ) of uncertain signals, e.g., prices and FR signals, to optimize the economic benefits of the battery system for simultaneously participating FR markets and demand-side management. \citeauthor{kumar2019benchmarking}\cite{kumar2019benchmarking} presented a stochastic MPC scheme to optimize the profit of participating FR markets, and provided a comprehensive comparison of control performance with deterministic and stochastic MPC. It should be pointed out that the above FR strategies only consider FR signals at low time resolution (only hourly averaged FR signals are considered in both strategy design and numerical simulation). However, the hourly averaged FR signals cannot reflect the actual dynamics of FR requirements. Another common limitation of the above FR strategies is that only low-fidelity battery models are considered. Although low-fidelity models have advantages in computational complexity, they fail to capture accurate battery dynamics, e.g., capacity fade, and hence fail to provide reliable evaluation of the proposed FR strategies. In lithium-ion batteries, which are extensively used in both consumer electronics and industry, an irreversible reaction occurring at the boundary of electrolyte and electrode can create a so-called solid electrolyte interphase (SEI) layer and lead to the decay of battery capacity \cite{santhanagopalan2006review,peled1979electrochemical,wang2004lithium}. Compared with low-fidelity models, physics-based models provide high-fidelity predictions of battery's internal states, which can reflect the battery degradation/capacity fade and the satisfaction of some safety constraints. Considering a high-fidelity battery model and high-resolution FR signals, \citeauthor{cao2020multiscale}\cite{cao2020multiscale} designed a nonlinear MPC approach for optimizing the incentives of FR participation and minimizing battery degradation.

It is worth pointing out that while the above MPC-based FR strategies have achieved satisfactory control performance, real-time implementation of MPC requires solving optimization problems online, which can be large-scale and highly non-convex, at each time step with updated information. This requirement limits the applicability of MPC-based FR schemes in practice. In addition, the incurred computational overhead can also 
reduce the actual economic benefits for participating FR markets. One promising solution to facilitate the online implementation of MPC and reduce the computational burden is approximating the original MPC law using function approximation techniques, e.g., artificial neural networks. In this way, the online implementation of the approximated MPC law only requires simple function evaluation and is computationally efficient. The DNN-approximated fast MPC design has been attracting increasing attention in control community \cite{zhang2020near,hertneck2018learning,maddalena2020neural,vaupel2020accelerating,cao2020deep}. However, to the best of our knowledge, fast MPC for designing FR strategies is still not well investigated in the existing literature.

Motivated by the above discussions, this paper presents a deep learning-based MPC framework for optimizing the economic benefits of FR participation of stationary battery systems. Supervised learning (SL) and reinforcement learning (RL) algorithms are combined to give a fast MPC design of FR strategy with reinforced performance and low online computational cost. Specifically, the MPC formulation embedding a low-fidelity battery model is firstly designed with synthetic signals used in the FR setting to provide demonstrations for subsequent learning processes. Then, DNNs are trained via SL algorithms using the given demonstrations to learn an approximation of the MPC policy with the low-fidelity model. We note that while the low-fidelity battery model is computationally inexpensive, it does not capture the effects of battery degradation. To address this issue, by applying the DNN-approximated policy to interact with a high-fidelity battery simulator, we adopt an RL algorithm, called deterministic policy gradient (DPG), to tune the DNN parameters so that the DNN-approximated FR policy can trade off between short term economic incentives and long term battery degradation to prolong the battery's lifetime and improve the economic benefits of FR participation.


\section{Preliminaries}
\subsection{Single Particle Battery Model}
A mathematical model of batteries is essential in analyzing battery properties and designing battery management schemes. In this section, we introduce a single particle (SP) model of lithium-ion batteries, which will be used in subsequent sections to construct a high-fidelity simulator for implementing our proposed control scheme. Although the SP model is less accurate than other full-order electrochemical models such as the Doyle-Fuller-Newman model, it is computationally inexpensive. The SP model can be described by the following nonlinear differential-algebraic equations (DAEs):
\begin{subequations}\label{eq:model1}
\begin{align}
&\frac{dc_i^{avg}}{dt} = \frac{-3J_j}{R_jF},\quad j\in\{n,p\},\\
&c_j^s = c_j^{avg}+\frac{-J_jR_j}{5D_jF},\quad j\in\{n,p\},\\
&J_j = 2\cdot i_{0,j}\cdot\sinh \left(\frac{0.5F}{RT}\eta_j\right),\quad j\in\{n,p\},\\
&i_{0,j} = Fk_j(c_{j,max}-c_j^s)^{0.5}{c_j^s}^{0.5}c_e^{0.5},\quad j\in\{n,p\},\\
& \eta_p = \phi_p - U_p(\theta_p),\\
& \eta_n = \phi_n - U_n(\theta_n)+\frac{R_fI_{app}}{S_n},\\
&\theta_j = \frac{c_j^s}{c_{j,\max}},\quad j\in\{n,p\}.
\end{align}
\end{subequations}

Following the results given by \citeauthor{cao2020multiscale}\cite{cao2020multiscale} and \citeauthor{moura2010optimal}\cite{moura2010optimal}, we assume a side reaction leading to the formation of a resistive SEI film in the negative electrode which causes the loss of active material and battery capacity fade. The side reaction and the SEI film growth are governed by the following equations:
\begin{subequations}\label{eq:model2}
\begin{align}
&J_{sd} = -i_{o,sd}\exp(-F\eta_{sd}/RT),\\
&\eta_{sd} = \phi_n - U_{ref}+\frac{R_fI_{app}}{S_n},\\
&R_f = R_{SEI}+ \delta_f/\kappa_{sd},\\
& \frac{d\delta_f}{dt} = \frac{-J_{sd}M_{sd}}{\rho_{sd}F},\\
& C^f = \int \frac{J_{sd}S_n}{Q_{\max}} dt.
\end{align}
\end{subequations}
The voltage $V$, current $I_{app}$, power $P$, and energy $E$ of the battery are computed as
\begin{subequations}\label{eq:model3}
\begin{align}
&V = \phi_{p} - \phi_n,\\
&J_p = \frac{I_{app}}{S_p},\\
&J_n+J_{sd} = \frac{-I_{app}}{S_n},\\
&P = I_{app}V,\\
&E = \frac{c_{n}^{avg}}{c_{n,\max}}\overline{E}.
\end{align}
\end{subequations}

\begin{table}
\caption{List of symbols for SP battery model}
\label{tb:symbols}
\centering
\resizebox{\textwidth}{!}{
\begin{tabular}{ll|ll}\toprule\hline
$D_j$ & \makecell[l]{solid surface diffusion coefficient of lithium in\\ the electrode $j$ (m$^2$/s)} & $E$ & battery SOC (MWh)\\\hline
$\overline{E}$ & battery capacity (MWh) & $F$ & Faraday constant (96,487 C/mol)\\\hline
$c_j^{avg}$ & \makecell[l]{average concentration within the particle in\\ the electrode $j$ (mol/m$^3$)} & $c_j^s$ & \makecell[l]{surface concentration of lithium in\\ the electrode $j$ (mol/m$^3$)}\\\hline
$c_e$ & \makecell[l]{concentration of electrolyte in solution phase (mol/m$^3$)}
& $C^f$ & capacity fade \\\hline
$i_{o,sd}$ & \makecell[l]{exchange current density for side reaction (A/m$^2$)} & $I_{app}$ & applied current passing through the cell (A)\\\hline
$j$ & negative $n$ and positive $p$ electrodes & $J_j$ & \makecell[l]{local reaction current density referred to the surface\\ area of electrode $j$ (A/m$^2$)} \\\hline
$k_j$ & rate constant of electrochemical reaction ($m^{2.5}/(mol^{0.5}s)$) & $M_{sd}$ & molecular weight of SEI (kg/m$^{3}$)\\\hline
$P$ & power supplied to the battery & $R_j$ & particle radius for electrode $j$ (m)\\\hline
$R_f$ & resistance of the SEI film ($\Omega\cdot m^2$) & $R_{SEI}$ & initial resistance of the SEI film ($\Omega\cdot m^2$) \\\hline
$sd$ & side reaction & $S_j$ & total electroactive surface area of electrode $j$ ($m^2$)\\\hline
$t$ & time & $T$ & temperature \\\hline
$U_j$ & local equilibrium potential of the side reaction ($V$) & $U_{ref}$ & \makecell{constant equilibrium potential of the side reaction ($V$)} \\\hline
$V$ & cell voltage ($V$) & $\phi_j$ & solid phase potential of electrode $j$ ($V$)\\\hline
$\kappa_{sd}$ & SEI ion conductivity ($S/M$) & $\eta_j$ & local overpotential of electrode $j$ ($V$) \\\hline
$\rho_{sd}$ & density of SEI ($kg/m^3$) & $\delta_{f}$ & SEI film thickness ($m$)\\\hline
\end{tabular}}
\end{table}
Symbols in \eqref{eq:model1}, \eqref{eq:model2} and \eqref{eq:model3} are explained in Table \ref{tb:symbols}. More details about the SP battery model can be found in the works by \citeauthor{santhanagopalan2006review} and \citeauthor{cao2020multiscale}, and references therein\cite{santhanagopalan2006review,cao2020multiscale}. The above SP model provides a high-fidelity prediction of battery states and captures the capacity fade incurred by charging/discharging cyclings. This SP model will be utilized in our paper to serve as a high-fidelity battery simulator for simulating and testifying our proposed FR policies. 
\subsection{Low-fidelity Battery Model}
While the above SP model can reflect the detailed states, embedding this model in the MPC design can be computationally expensive in that a large-scale nonconvex optimization problem needs to be solved in real-time. To make the MPC formulation computationally tractable, we introduce the following low-fidelity battery model:
\begin{equation}\label{eq:low_fidelity1}
E_{t+1} = E_{t} + \Delta t\cdot P_{t},
\end{equation}
where $E_t$ is the battery SOC at $t$-th time step, $\Delta t$ is the sampling period, and $P_t$ denotes the charging/discharging rate at $t$-th time step. This low-fidelity model is solely based on energy balance and is widely used in the existing literature for the reason of computational simplicity. The limitation of this model is that it cannot reflect battery degradation. Consequently, the MPC design adopting this model is incapable of explicitly considering the battery capacity fade.
\section{FR Market Participation Problem}

\subsection{Decision-making Setting}
\begin{figure}
  \centering
  \includegraphics[width=0.8\textwidth]{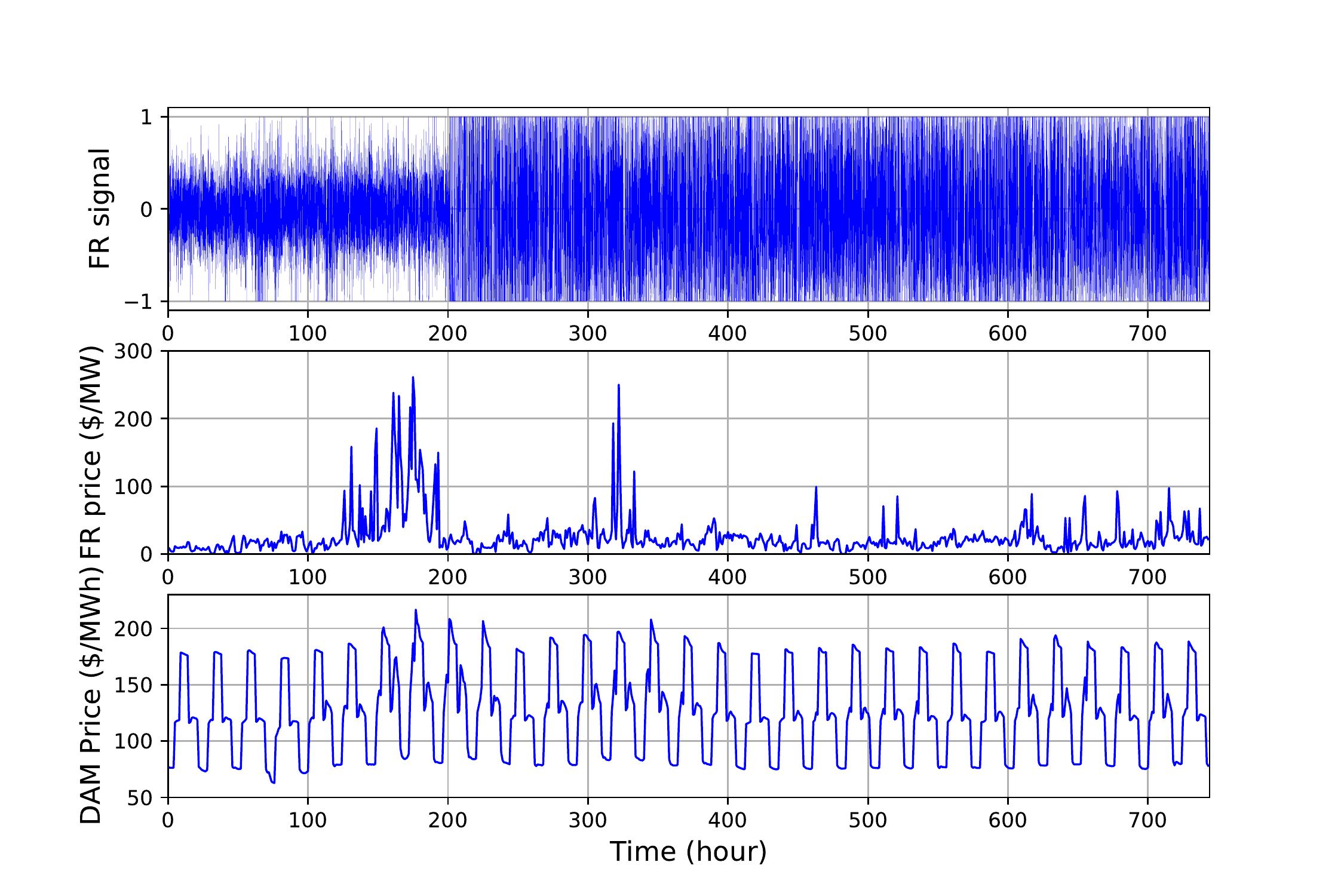}\\
  \caption{PJM data for one month of FR signals, FR capacity prices, and DAM electricity prices.}
  \label{demo}
\end{figure}

We begin by illustrating the decision-making setting of FR participation with stationary battery systems. In our paper, we consider designing real-time optimal participation strategies for the battery system in FR market that is operated by ISO. The various cost and revenue components considered in our decision-making setting are given as follows:
\begin{itemize}
\item FR capacity (hourly): Each hour, the battery system determines the FR capacity committed to the grid for the next immediate hour. The committed FR capacity remains constant during each hour. Given the FR band, the ISO can request a fraction of the committed capacity according to real-time grid requirements by using FR signals, which vary every 2 seconds and have a range of $[-1,1]$. The ISO remunerates the battery for dispatching the FR capacity according to hourly time-varying FR prices.
\item Power Purchase (hourly): The battery system can purchase power from the grid to replenish itself. The price of purchasing power from the grid varies hourly according to a day-ahead energy market (DAM).
\item Load (hourly): An adjustable load can draw power from the battery system to help adjust the remaining energy of the battery system. This load is updated hourly and maintained as constant during the whole hour. The cost of this load is assumed to be zero.
\end{itemize}
Fig. \ref{demo} depicts the real data for one month of FR signals, FR capacity prices, and electricity prices from PJM.

\subsection{MPC Formulation Framework}
The MPC formulation for optimizing FR participation of the battery system has multi-scale characterization in that it needs to optimize short-term economic incentive, which is determined by electricity prices, FR prices, and FR signals while capturing the effects of long-term battery degradation. Under the MPC framework, an optimization problem is solved every $t$-th hour for a set of signals over a finite prediction horizon $\mathcal{N}_t:=\{t+1,t+2,\cdots,t+N\}$, where $N$ is the length of the prediction horizon. Since FR signals are updated every 2 seconds, we discretize every hour with $S = 1800$ steps and define a time interval set $\mathcal{S} := \{1,2,3,\cdots,S\}$.

For our considered MPC formulation, the parameters, data, and decision variables are summarized as below:
\begin{itemize}
\item Parameters and data: $\pi_t^e\in\mathbb{R}_+$ denotes the electricity price at $t$-th hour, $\pi_t^f\in\mathbb{R}_+$ represents the FR price at $t$-th hour, $\alpha_{t,s}\in[-1,1]$ is the FR signal denoting the fraction of FR capacity requested by the ISO at $t$-th hour and $s$-th step ($\alpha_{t,s}>0$ means the ISO sends power to the battery, and $\alpha_{t,s}<0$ means the ISO withdraws power), $\overline{P}\in\mathbb{R}_+$ and $\underline{P}\in\mathbb{R}_+$ are the maximum charging rate and discharging rate, respectively, $\overline{E}\in\mathbb{R}_+$ is the battery capacity.
\item Variables: $F_t\in\mathbb{R}_+$ is the committed FR capacity provided by the battery system, $P_{t,s}$ denotes the charging/discharing rate at $t$-th hour and $s$-th step ($P_{t,s}>0$ means the battery is being charged and $P_{t,s}<0$ means the battery is being discharged), $O_t$ is the power purchased from the grid at $t$-th hour, $E_{t,s}\in\mathbb{R}$ is the battery state of charge (SOC) at $t$-th hour and $s$-th step.
\end{itemize}
All above quantities with single subindex $t$, namely ($\pi_t^e$, $\pi_t^f$, $F_t$, $O_t$ and $L_t$), vary hourly and remain constant over the time interval $[t-1,t]$. Quantities with subindexes $t$ and $s$, namely $(\alpha_{t,s},E_{t,s},P_{t,s})$, are updated every 2 seconds and are held constant over time interval $[t+\frac{s-1}{S},t+\frac{s}{S}]$.


%

\subsubsection{Objective Function}
The design objective is to maximize the net profit of FR participation for the battery system by optimizing the charging/discharging action of the battery system, the committed FR capacity to the ISO, and the power purchased from the power grid. For the MPC formulation, we maximize the following objective function at every $t$-th hour over the prediction horizon $\mathcal{N}_t$:
\begin{equation}\label{eq:obj}
\sum_{k\in\mathcal{N}_t}{\pi}_k^fF_k - \sum_{k\in\mathcal{N}_t}{\pi}_k^eO_k,
\end{equation}
where the first term denotes the revenue obtained from the provision of FR capacity, the second term represents the cost of buying energy from the grid. While the profit of FR participation is also dependent on the battery's lifetime, the MPC design embedding the low-fidelity model \eqref{eq:low_fidelity1} fails to consider battery's capacity fade. This issue will be solved in the subsequent RL-based design, which learns the effects of battery aging caused by FR participation from the data generated by interacting with the high-fidelity battery simulator.

\subsubsection{Constraints}
According to \eqref{eq:low_fidelity1}, the battery SOC dynamics embeded in the MPC design are given as 
\begin{equation}\label{eq:low_fidelity2}
E_{k,s+1} = E_{k,s} + \frac{1}{S}\cdot P_{k,s},\; k\in\mathcal{N}_t,\; s\in\mathcal{S},
\end{equation}
where the net charged/discharged power of batteries should be within the following admissible range 
\begin{equation}\label{eq:}
\underline{P}\leq P_{k,s}\leq \overline{P},\; k\in\mathcal{N}_t,\; s\in\mathcal{S}.
\end{equation}
In addition, at each time instant, the net charged/discharged power is equal to the amount of power transacted with the ISO due to FR participation plus the amount of power purchased from the grid minus the amount of the adjustable load
\begin{equation}
P_{k,s} = \alpha_{k,s}\cdot F_k + O_k - L_k, \; k\in\mathcal{N}_t,\; s\in\mathcal{S}.
\end{equation}
Within our design framework, we assume that the battery system does not simultaneously buy power from the grid and discharge to the load. As a result, $O_t$ and $L_t$ satisfy $O_t\cdot L_t =0$.

In order to reserve sufficient power to counteract system uncertainties and avoid fully charging and discharging the battery, the following constraints are imposed
\begin{eqnarray}
&\lambda_l(1-C_{k,s}^f)\overline{E} \leq E_{k,s} \leq \lambda_u(1-C_{k,s}^f)\overline{E},\quad k\in\mathcal{N}_t,\; s\in\mathcal{S},\\
&\epsilon_l(1-C_{t,1}^f)\overline{E} \leq E_{t+N-1,S} \leq \epsilon_u(1-C_{t,1}^f)\overline{E},
\end{eqnarray}
where $\overline{E}$ is the battery capacity, parameters $\lambda_l\in(0,1)$ and $\lambda_u\in(0,1)$ provide a safe margin for the battery SOC, $\epsilon_l\in(0,1)$ and $\epsilon_u\in(0,1)$ impose a terminal constraint on the remaining energy at the end of the prediction horizon to prepare the battery for the FR participation in next horizon. This paper assumes that the battery capacity fade $C^f$ is measurable. To keep the variables within reasonable ranges, we add the following simple bounds:
\begin{equation}\label{eq:FO_constraints}
0\leq F_{k} \leq \overline{P}, \quad 0\leq O_k\leq \overline{P},\quad k\in\mathcal{N}_t.
\end{equation}

\subsubsection{Low-fidelity MPC Strategy}
The MPC-based FR strategy aims at determining the optimal commitments $F_t$, $O_t$ and $L_t$ at each hour by maximizing the objective function \eqref{eq:obj} while guaranteeing constraints \eqref{eq:low_fidelity2}-\eqref{eq:FO_constraints} and the satisfaction of battery models. Since the low-fidelity battery model is used, we call this design low-fidelity MPC (LF-MPC). At each hour, the LF-MPC problem is solved to yield the optimal commitment vectors $F_{\mathcal{N}_t}, O_{\mathcal{N}_t}$ and $L_{\mathcal{N}_t}$. Then, only the next hour commitment decisions ($F_t$, $O_t$, $L_t$) are implemented and the prediction horizon $\mathcal{N}_t$ is shifted to next hour due to the receding horizon nature of MPC. It is worth remarking that there are several shortcomings of the LF-MPC. On the one hand, the LF-MPC is incapable of considering the battery capacity fade because the low-fidelity battery model \eqref{eq:low_fidelity2} does not reflect the evolution of battery degradation, and consequently, the computed FR commitment decisions can be aggressive and myopic. On the other hand, the implementation of this LF-MPC entails solving an optimization problem at each hour, which can give rise to high computational overhead and reduce the actual economic gain of FR participation.

\section{Learning-based FR strategies with Reinforced Approximation}

Motivated by the shortcomings of the LF-MPC, in this section we provide a learning-based FR strategy, which consists of two main parts that are conducted sequentially. Firstly, DNNs are trained via SL algorithms to yield an approximation of the LF-MPC policy and an actor-value function using the data generated by simulating the LF-MPC. Then, the RL algorithm, DPG, is deployed to finely adjust the parameters of the DNN-approximated policy to balance the short-term economic incentives and long-term battery degradation so that the battery's lifetime is prolonged and the economic benefit of FR participation is improved.

\subsection{Uncertainty Qualification}
During offline learning processes, we assume that the real data of the prices and the FR signals $(\pi_t^f,\pi_t^e,\alpha_{t,s})$ are unavailable, and only artificially generated uncertainty representations can be used. In our design, those uncertain signals are learned using historical data. Then, representations of these uncertain signals, denoted as ($\hat{\pi}_t^f,\hat{\pi}_t^e,\hat{\alpha}_{t,s}$), are utilized to replace the real data ($\pi_t^f,\pi_t^e,\alpha_{t,s}$) in offline learning procedures.

The uncertainty qualification techniques used in our paper are adopted from the work of \citeauthor{kumar2018stochastic}\cite{kumar2018stochastic}. In summary, the energy price is assumed to follow a weekly pattern, and are modelled as multi-dimensional Gaussian variables $\pi_T^e\sim \mathcal{G}(\overline{\pi}_T^e, \Sigma_{T}^e)$, where $\pi_T^e\in\mathbb{R}^{|T|}$ is the energy price vector for one week, $\overline{\pi}_T^e\in\mathbb{R}^{|T|}$ and $\Sigma_{T}^e\in\mathbb{R}^{|T|\times|T|}$ are corresponding mean vector and covariance matrix, respectively, ${T} := \{1,2,3,\cdots,168\}$ is the time index set for a whole week; temporal correlation patterns in $\pi_T^e$ are captured in the covariance matrix $\Sigma_{T}^e$. The FR price $\pi_t^f$ is assumed to follow a log-normal distribution $\pi_t^f\sim \mathcal{LN}\left(\ln(\overline{\pi}_t^f),\sigma_{\pi^f}^2\right)$. As for the FR signal vector $\alpha_{t,\mathcal{S}}:=[\alpha_{t,1},\cdots,\alpha_{t,S}]$ at every hour, since it is highly random and difficult to be characterized, we use randomly selected historical data as its forecast. Please refer to the work of \citeauthor{kumar2018stochastic}\cite{kumar2018stochastic} for explicit procedures of modeling these random variables.
\subsection{Supervised Learning for Approximating the LF-MPC}
To implement the LF-MPC, an optimization problem needs to be solved in real-time with updated information. Although MPC usually achieves satisfactory control performance and guarantees constraints satisfaction, its online computational cost can be pretty high. In order to facilitate the online implementation of the MPC-based FR strategy and reduce its online computation overhead, 
supervised learning is deployed in this paper to learn a DNN-based approximation of the LF-MPC policy.

A general framework for approximating a policy/control law is to firstly generate a set of state-control pairs $D := \{(x_1,a^*_1),\cdot,(x_N,a^*_N)\}$, where $x_i$ is the state vector and $a_i^*$ is the control input vector generated by a ``good'' control policy $\kappa(x_i)$ for $x_i$. In this paper, the optimal input $a^*_i$ is generated by solving the LF-MPC problem with $x_i$ as the characterization of necessary signals and parameters used in the LF-MPC. Then, for a parametric family of policies $\mu_{\theta}(x)$, its parameters $\theta$ are optimized to provide an approximation of $\kappa(x_i)$ by minimizing
\begin{equation}\label{eq:optimal}
\min_{\theta}\frac{1}{N}\sum_{i=1}^N L_{\theta}\left(a_i^* -\mu_{\theta}(x_i)\right),
\end{equation}
where $L_{\theta}(a_i^*-\mu_{\theta}(x_i))$ is a loss function penalizing the approximation error $a_i^*-\mu_{\theta}(x_i)$. By solving \eqref{eq:optimal}, the derived parametric policy $\mu_{\theta}(\cdot)$ usually is a good approximation of the original policy $\kappa(\cdot)$ if sufficient training data are provided.

A common choice of the parametric policy is a DNN, which is expressed as a function composition
\begin{equation}
\mu_{\theta}(x) := A_L\circ\lambda_L\cdots A_1\circ\lambda_1(x),
\end{equation}
where $\lambda_i(\xi) := W_i \xi + b_i$ are affine functions; $A_i$ are activations functions which generally are nonlinear. The parameters of the DNN $\mu_{\theta}(x)$ are $\theta:=\{W_1,b_1,\cdots,W_L,b_L\}$. 

In the SL process, uncertainty representations ($\hat{\pi}_{\mathcal{N}_t}^f, \hat{\pi}_{\mathcal{N}_t}^e, \hat\alpha_{\mathcal{N}_t}$) are deployed when implementing the LF-MPC. Solving the LF-MPC at $t$-th hour requires samples of uncertain parameters $ (\hat{\pi}_{\mathcal{N}_t}^f, \hat{\pi}_{\mathcal{N}_t}^e,\hat{\alpha}_{\mathcal{N}_t})$, battery SOC $E_{t,1}$ and capacity fade $C^f_{t,1}$. Since $\hat{\alpha}_{\mathcal{N}_t}$ is with high time resolution and large dimension, in order to make the size of the DNN structure within reasonable range, we take as DNN input $x_t := [\text{mean}(\hat{\alpha}_{\mathcal{N}_t}),\text{var}(\hat{\alpha}_{\mathcal{N}_t}),\hat{\pi}^f_{\mathcal{N}_t},\hat{\pi}_{\mathcal{N}_t}^e,E_{t,1},C^f_{t,1}]^{\mathrm{T}}$, where $\text{mean}(\cdot)$ and $\text{var}(\cdot)$ denote the mean and the variance of the associated FR signal vector, respectively. Since it is assumed that the battery does not buy power from the grid while discharging to the load, which means $O_t\cdot L_t = 0$, so $O_t$ and $L_t$ can be exclusively determined by the value of $O_t - L_t$. Hence, we select the target inputs $a^*_t$ for $x_t$ as $a_t^* := [F_t,O_t-L_t]^{\mathrm{T}}$ that are computed via the LF-MPC. 

While a well-trained DNN generally provides a good approximation of the original policy, it suffers from several shortcomings under our design framework. On the one hand, the LF-MPC formulation only considers the low-fidelity battery model, which does not reflect the battery degradation. On the other hand, only a short prediction horizon $\mathcal{N}_t$ can be considered in the LF-MPC for making the resultant optimization problem computationally tractable. Consequently, the FR policy obtained via the LF-MPC can be aggressive and myopic. In addition, approximation error of neural networks can also degrade the performance of the DNN-approximated policy. All the above issues motivate the subsequent RL-based design to improve the performance of the DNN-approximated policy utilizing RL algorithms.

\subsection{Reinforcement Learning via Deterministic Policy Gradient}
Recall that the LF-MPC is incapable of considering the battery degradation since the low-fidelity model does not to reflect the dynamics of capacity fade. Hence, the LF-MPC, as well as the DNN-approximated FR policy derived in the above subsection, can be aggressive and myopic. In addition, while the high-fidelity SP model does capture the effects of battery aging, embedding it into the MPC design gives rise to extremely high online computational cost, and the robustness of the corresponding MPC policy is subject to the solvability of resultant nonlinear optimization problems. Hence, how to design an FR strategy to deal with the battery degradation while maintaining low online implementation cost is worth investigating. 

This subsection provides a RL algorithm, called deterministic policy gradient (DPG), to adjust the DNN-approximated policy to mitigate the battery aging and improve the economic benefit of FR participation. RL offers systematic tools for designing control policies for dynamic systems without having the exact system models and distributions of system uncertainties\cite{sutton2018reinforcement,bertsekas2019reinforcement}. In the context of RL, for a dynamic system $x_{t+1} = f(x_t,a_t)$, where $x_t$ is system states vector, $a_t$ control inputs vector, and $f(x_t,a_t)$ transition dynamics, a stage reward $r(x_t,a_t,x_{t+1})\in\mathbb{R}$ will be received for the transition $x_t\stackrel{a_t}{\longrightarrow}x_{t+1}$. By letting the controller interact with the dynamic system, the DPG algorithm aims at optimizing a parametric policy $\mu_{\theta}(\cdot)$ to maximize the following cumulative rewards over infinite horizon:
\begin{equation}\label{eq:return}
J(\pi_{\theta}) = \mathbb{E}\left[\sum_{k = 0}^{\infty}\gamma^k\cdot r(x_k,a_k,x_{k+1})\mid_{a_k = \pi_\theta(x_k)}\right],
\end{equation}
where the discount factor $\gamma$ is used to weight the immediate reward versus long term rewards, the expected value $\mathbb{E}(\cdot)$ is defined over the probability measure in both transition dynamics and stage rewards. 

For our design framework, we can model the decision-making process of FR as the following dynamic system
\begin{equation}
x_{t+1} = f(x_t,a_t)
\end{equation}
where $x_t:=[\text{mean}(\alpha_{t,\mathcal{S}}),\text{var}(\alpha_{t,\mathcal{S}}),\pi^f_{t},\pi^e_{t},E_{t,1},C_{t,1}^f]^{\mathrm{T}}$ and $a_t:=[F_t,O_t-L_t]^{\mathrm{T}}$. It is clear that this dynamic system is stochastic and uncertain in that the transition dynamics are dependent on the high-fidelity battery model and the uncertain signals used for FR. Fortunately, the DPG algorithm is model free and does not rely on exact system dynamics. It can learn from the data of transition tuples  $(x_t, a_t, r_t,x_{t+1})$ to maximize the objective function \eqref{eq:return}.

The stage reward considered in the RL process is defined as
\begin{equation}\label{eq:obj_rl}
r(x_t,a_t,x_{t+1}):={\pi}_t^fF_t - {\pi}_t^eO_t-\pi^{C^f}\left(C^f_{t+1,1} - C^f_{t,1}\right),
\end{equation}
where the first term denotes the revenue of the committed FR capacity, the second term is the cost of buying power from the grid, the last term penalizes the capacity fade incurred by charging/discharging the battery. In comparison with the objective function considered in the LF-MPC design in \eqref{eq:obj}, an extra penalty term of battery degradation is considered in \eqref{eq:obj_rl}. The reason for the inclusion of this term is motivated by the fact that the battery's lifetime can be decreased by frequent charging/discharging cycling for FR participation. We assume that the battery reaches its end of life (EOL) if its capacity fade is over a certain threshold. Consequently, it is profitable for the battery system to prolong its service time by limiting FR commitments to some extent. It is worth remarking here that the last term in \eqref{eq:obj_rl} cannot be handled in the LF-MPC formulation since the low-fidelity battery model \eqref{eq:low_fidelity2} cannot reflect the evolution of capacity fade $C^f$ . In addition, the objective function in \eqref{eq:obj_rl} optimizes the cumulative rewards over an infinite horizon. 
In contrast, the LF-MPC only considers a short prediction horizon $\mathcal{N}_t$ for the reason of computational tractability. Consequently, it can be expected that the FR policy derived via RL-based methods in this subsection is more strategic and far-sighted than the LF-MPC.

A frequently used concept in many RL algorithms is the action-value function $Q^{\pi}(x_t,a_t)$ corresponding to a given policy $\pi$, which is defined as
\begin{equation}
Q^{\pi}(x_t,a_t) := \mathbb{E}\left[\sum_{k=t}^\infty\gamma^{k-t}\cdot r(x_k,a_k,x_{k+1})\mid x_t,a_t\right].
\end{equation}
It denotes the expectation of the discounted return of taking action $a_t$ in state $x_t$ and thereafter following the policy $\pi$. According to the DPG method developed by \citeauthor{silver2014deterministic}\cite{silver2014deterministic}, the gradient of $J(\pi_{\theta})$ in \eqref{eq:return} w.r.t. $\theta$ is
\begin{equation}\label{eq:q}
\nabla_{\theta} J(\pi_{\theta}) = \mathbb{E}\left[\nabla_{\theta}\pi_{\theta}(x)\nabla_{a}Q^{\pi_{\theta}}(x,a)\mid _{a = \pi_{\theta}(x)}\right].
\end{equation}
By replacing the action-value function $Q^{\pi_{\theta}}(x,a)$ in \eqref{eq:q} with a parametric approximator $Q_w(x,a)$ parameterized by $w$, it gives an approximation of $\nabla_{\theta}J(\pi_{\theta})$ as
\begin{equation}\label{eq:gradient}
\nabla_{\theta}J(\pi_{\theta}) \approx \mathbb{E}\left[\nabla_{\theta}\pi_{\theta}(x)\nabla_a Q_w(x,a)\mid_{a = \pi_{\theta}(x)}\right].
\end{equation}
With the gradient information in \eqref{eq:gradient}, the objective function $J(\pi_{\theta})$ can be optimized by updating the parameter $\theta$ using gradient-based optimization algorithms.

The action-value function $Q^{\pi}(x,a)$ has the following recursive relationship, known as Bellman equation:
\begin{equation}
Q^{\pi}(x_t,a_t) = \mathbb{E}\left[r(x_t,a_t,x_{t+1}) + \gamma Q^{\pi}(x_{t+1},\pi(x_{t+1}))\right].
\end{equation}
Consequently, given a batch of $N$ transitions $(x_i,a_i,r_i,x_{i+1},a_{i+1})$, we can train the action-value function approximator $Q_w(x,a)$ by minimizing the following objective function
\begin{equation}\label{eq:tderror}
\min_{w}\frac{1}{N}\sum_{i=1}^NL_{w}\left(Q_w(x_i,a_i)-y_i\right),
\end{equation}
where $y_i = r_i + Q_w(x_{i+1},a_{i+1})$, the loss function $L_w$ penalizes the temporal difference (TD) error $Q_w(x_i,a_i) - y_i$.

Directly utilizing RL algorithms to learn a FR policy from scratch can be data-intensive and time-consuming, and can cause unnecessary wear and tear for physical systems when doing hardware-in-the-loop simulations in that the performance of RL algorithms in the early stage of learning process is usually poor. As a result, we exploit the data generated via the LF-MPC to learn proper initializations for both the FR policy $\mu_{\theta}(\cdot)$ and the action-value function $Q_w(\cdot,\cdot)$ used in the RL process. Given the transitions  $(x_t,a^*_t,r_t,x_{t+1},a^*_{t+1})$ generated via the LF-MPC, where $x_t := [\text{mean}(\hat{\alpha}_{t,\mathcal{S}}),\text{var}(\hat{\alpha}_{t,\mathcal{S}}),\hat{\pi}^f_t,\hat{\pi}_t^e,E_{t,1},C^f_{t,1}]^{\mathrm{T}}$,
 $a^*_t := [F_t,O_t-L_t]^{\mathrm{T}}$, and $r_t := \pi_{t}^{f}F_t-\pi_t^eO_t-\pi^{C^f}(C_{t+1,1}^f - C_{t,1}^f)$, a DNN-approximated policy $\mu_{\hat\theta}(\cdot)$ and an action-value function $Q_{\hat{w}}(\cdot,\cdot)$ can be learned by minimizing the loss functions in \eqref{eq:optimal} and \eqref{eq:tderror}, respectively. Then, $\mu_{\hat{\theta}}(\cdot)$ and $Q_{\hat{w}}(\cdot,\cdot)$ will be utilized to initialize the actor and the critic in the subsequent RL process. 
 
\subsection{Implementation}
The proposed machine learning-based predictive control scheme for battery management comprises two main parts: SL process and RL process. In the SL process, the LF-MPC is simulated firstly to generate a number of transition tuples $(x_t,a^*_t,r_t,x_{t+1},a^*_{t+1})$. Then, neural networks are trained by minimizing the loss functions in \eqref{eq:optimal} and \eqref{eq:tderror} to give a policy $\mu_{\hat\theta}(\cdot)$ and an action-value function $Q_{\hat{w}}(\cdot,\cdot)$. In the RL process, by initializing the policy and the actor-value function with $\mu_{\hat{\theta}}(\cdot)$ and $Q_{\hat{w}}(\cdot,\cdot)$ that are derived in the SL process, respectively, the DPG algorithm is then deployed to adapt the DNN-approximated FR policy to balance economic incentives and battery degradation. In our work, the RL process is continued until a brand new battery reaches its EOL. As suggested by \citeauthor{lillicrap2015continuous}\cite{lillicrap2015continuous}, in the RL process, we deploy ``replay memory'' and ``target network'' with ``soft target updates'' to make the learning process more stable and efficient. In addition, since the electricity price is assumed to follow a weekly pattern, we select one week as one episode, and episodically update $\mu_{\theta}(\cdot)$ and $Q_{w}(\cdot,\cdot)$ as well as their target networks using gradient-based methods for $T$ iterations. During each episode, the policy $\mu_{\theta}(\cdot)$ 
interacts with the high-fidelity battery simulator to generate transition tuples $(x_t,a_t,r_t,x_{t+1})$ to update the replay memory buffer. Detailed procedures of the proposed scheme are summarized in Algorithm 1. 

\begin{algorithm}[H]
\label{alg:1}
\caption{Learning-based predictive control of battery management for FR}
\begin{algorithmic}[1]
\Statex \textit{Supervised Learning Process}
\vspace{2pt}
\State Generate synthetic signals $(\hat{\alpha}_{\mathcal{N}_t}, \hat{\pi}_{\mathcal{N}_t}^f,\hat{\pi}_{\mathcal{N}_t}^e)$ using UQ procedures
\State Simulate the LF-MPC with synthetic signals on the SP model to generate transitions $(x_t,a_t^*,r_t,x_{t+1},a^*_{t+1})$
\State Choose DNN structures for the FR policy $\mu_{\theta}(\cdot)$ and the action-value function $Q_{w}(\cdot,\cdot)$
\State Learn a DNN-approximated policy $\mu_{\hat\theta}$ by minimizing the loss function \eqref{eq:optimal}
\State Learn an action-value function approximator $Q_{\hat w}$ by minimizing the loss function \eqref{eq:tderror}
\end{algorithmic}
\vspace{2pt}

\begin{algorithmic}[1]
\Statex\textit{Reinforcement Learning Process}
\vspace{2pt}
\State{Set the battery SOC corresponding to half its capacity and $C^f = 0$}
\State Initialize the actor $\mu_{\theta}(\cdot)$ and the critic $Q_w(\cdot,\cdot)$ with weights $\theta \leftarrow\hat{\theta}$ and $w\leftarrow\hat{w}$
\State Initialize target networks $\mu_{\theta'}'(\cdot)$ and $Q_{w'}'(\cdot,\cdot)$ with weights $\theta'\leftarrow\hat{\theta}$ and $w'\leftarrow\hat{w}$
\While{$C^f\leq 0.2$}
\State Initialize a random process $\mathcal{N}$ for action exploration
\State Simulate the policy $a_t = \mu_{\theta}(x_t)+\mathcal{N}$ on the SP model for one episode using synthetic data $(\hat{\alpha}, \hat{\pi}^f,\hat{\pi}^e)$, and restore all transitions $(x_t,a_t,r_t,x_{t+1})$ in the replay memory ($RM$)
\For{$k$ = 1 to $T$}
\State Sample a batch of $N$ transitions $(x_i,a_i,r_i,x_{i+1})$ from the $RM$
\State Set $y_i = r_i + Q'(x_{i+1},\mu_{\theta'}(x_{i+1}))$, and update the critic network $Q_w(\cdot,\cdot)$ by minimizing the loss function in \eqref{eq:tderror}
\State Update the policy network $\mu_{\theta}(\cdot)$ using the sampled gradient information:
\Statex \begin{center}$\nabla_{\theta}J\approx\frac{1}{N}\sum_{i}\nabla_{\theta}\mu_{\theta}(x_i)\nabla_{a}Q_{w}(x,a)\mid_{x=x_i,a=\mu_{\theta}(x_i)}$
\end{center}
\State Update the critic network $Q_w(\cdot,\cdot)$ by optimizing the loss: $\frac{1}{N}\sum_{i}L_w(Q_w(x_i,a_i) - y_i)$
\State Update the target networks using ``soft target updates'':
\Statex \begin{center}$\theta' = \tau \theta + (1-\tau)\theta',\quad w'=\tau w+(1-\tau)w'$
\end{center}
\EndFor
\EndWhile
\end{algorithmic}
\end{algorithm}

In this paper, it is assumed that the battery reaches its EOL if the capacity fade $C^f \geq 20\%$. We compare the performance of different FR strategies according to their net profit of FR participation earned before battery's EOL:
\begin{equation}
\sum_{t=0}^{EOL}\pi_t^fF_t - \pi_t^eO_t.
\end{equation}
The online implementation procedures of FR strategies are summarized in Algorithm 2.
\begin{algorithm}[H]
\caption{Implementation of FR strategies}
\begin{algorithmic}[1]\label{algo2}
\State STRAT $t = 0$ with battery SOC corresponding to half of its capacity and the capacity fade $C_{0,1}^f = 0$.
\State Implement FR strategies using real data $(\alpha_{t,\mathcal{S}},\pi_{t}^f,\pi_{t}^e)$ to obtain FR commitments $F_{t+1}$, $O_{t+1}$ and $L_{t+1}$.
\State Inject decisions over $(t,t+1)$, calculate the battery net charging/discharging rate $P_{t+1,s} = \alpha_{t+1,s}\cdot F_{t+1} +O_{t+1} -L_{t+1}$.
\State Simulate the high-fidelity battery model with $P_{t+1,s}$ to update battery internal states and get $C_{t+1,S}^f$.
\State If $C^f_{t+1,S}\geq 0.2$, set $EOL = t$ and BREAK.
\State Set $t\leftarrow t+1$ and return to step 2.
\end{algorithmic}
\end{algorithm}

\section{Simulation Results}
We consider a battery comprised A123 system and ANR26650M1 cells with lithium-ion phosphate cathodes. The parameters for each cell are adopted from the work by \citeauthor{forman2012genetic}\cite{forman2012genetic}. The number of cells is scaled so that the total capacity is equal to 1 MW. The maximum charging/discharging rate is set as 10 MW. To accurately capture the FR signals, we discretize battery models with timesteps of 2 seconds, resulting in 1800 steps per hour. For the LF-MPC design, we set the prediction horizon as one hour $N = 1$, $\lambda_l=0.1$, $\lambda_h = 0.9$, and $\epsilon_l = \epsilon_h = 0.5$. The optimization problem of the LF-MPC is an LP with 3606 variables. This optimization problem is modeled using {\tt Julia} package {\tt JuMP} and solved via {\tt GLPK} solver. Other remaining optimization problems in our simulation are solved using {\tt IPOPT}.  

The DNN structure of the parametric policy $\mu_{\theta}(\cdot)$ has 5 layers: 6 neurons in the input layer, $30-15-2$ neurons in subsequent hidden layers, and 2 neurons in the output layer. Activation functions in all hidden layers are selected as {\tt ReLU} except that the activation functions in the last hidden layer are selected as {\tt tanh}. To guarantee input constraints satisfaction, the output layer projects values from $[-1,1]$ to allowed ranges. The DNN structure for the action-value function $Q_w(\cdot,\cdot)$ has 4 layers with structure $8-30-15-1$, and all activation functions are selected as {\tt ReLU}. All DNNs in our simulation are modelled using the {\tt Julia} package {\tt Flux}. The loss functions in \eqref{eq:optimal} and \eqref{eq:tderror} are selected as mean absolute error (MAE) (namely $L_{\theta}(x) = L_w(x) = |x|_1$), and are optimized via {\tt ADAM}.

Historical data covering a whole year of FR signals, FR prices, and electricity prices from PJM are used for constructing uncertainty representations using UQ procedures. In the SL process, the LF-MPC is simulated with synthetic signals $(\hat{\alpha}_{\mathcal{N}_t},\hat{\pi}_{\mathcal{N}_t}^e,\hat{\pi}_{\mathcal{N}_t}^f)$ for the time period of 10 battery's EOL to generate transition tuples $(x_t,a_t,r_t,x_{t+1},a_{t+1})$. Then, $\pi_{\hat\theta}(\cdot)$ and $Q_{\hat{w}}(\cdot,\cdot)$ are trained using the generated transitions by minimizing the objective functions \eqref{eq:optimal} and \eqref{eq:tderror} with 5000 and 2000 epochs, respectively.

For the  RL process, we select $\tau = 0.01$ and $\gamma = 0.9$. Random samples of a Gaussian distribution $\mathcal{G}(0,0.0025)$ are injected in the output of the last hidden layer of the actor network $\mu_{\theta}(\cdot)$ to add noise for action exploration. The RM buffer has a size of 1680 tuples $(x_t,a_t,r_t,x_{t+1})$. The stage cost function at each hour is set as
\begin{equation}
r(x_t,a_t,x_{t+1}):=\pi_t^fF_t  - \pi_t^eO_t - \pi^{C^f}\left(C_{t+1,1}^f - C_{t,1}^f\right)
-5\left(E_{t+1,1}/\overline{E} - 0.5(1-C^f_{t,1})\right)^2,
\end{equation}
where the last term is to force the battery SOC at the end of each hour close to half of the remaining capacity for the preparation of next hour FR participation. The parameter $\pi^{C^f}$, which represents the long-term valuation of battery capacity, is set as $\$ 12,000$ based on the study of \citeauthor{cao2020multiscale}\cite{cao2020multiscale}. Since the electricity prices follow a weekly pattern, we select one week as one episode. After interacting with the high-fidelity battery simulator for one episode, parameters of the policy $\mu_{\theta}(\cdot)$ and the actor-value function $Q_w(\cdot,\cdot)$ are updated for $168\times4$ iterations using a batch of $160$ transitions sampled from the RM buffer.

In our simulation, we compare the following FR strategies:
\begin{itemize}
\item LF-MPC: the FR policy derived by solving the LF-MPC problem.
\item SL: the DNN-approximated policy trained only via SL.
\item SL \& RL: the DNN-approximated policy trained via both SL and RL.
\item RL: the DNN-approximated policy trained only via RL.
\item HF-MPC: the FR policy derived by solving the MPC embedding the high-fidelity SP battery model (proposed by \citeauthor{cao2020multiscale}\cite{cao2020multiscale}).
\end{itemize}
Due to the discretization error of the battery model, the adoption of low-fidelity model in the LF-MPC design, and the inability of RL algorithms in dealing with hard constraints, it is possible that the market commitment decisions $(F_{t+1},O_{t+1},L_{t+1})$ obtained via above FR strategies are infeasible. In these cases, to avoid fully charging/discharging the battery, we adjust the FR band as $F_{t+1}\leftarrow F_{t+1}-\Delta F$, where $\Delta F = 0.5MW$, and then recalculate the values of $O_{t+1}$ and $L_{t+1}$. All computations in our simulation were performed on the HPC cluster {\tt ComputeCanada}. Scripts for reproducing our simulation results are available in \url{https://github.com/li-yun/battery_FR_RL}.

\begin{figure}
  \centering
  \includegraphics[width=0.7\textwidth]{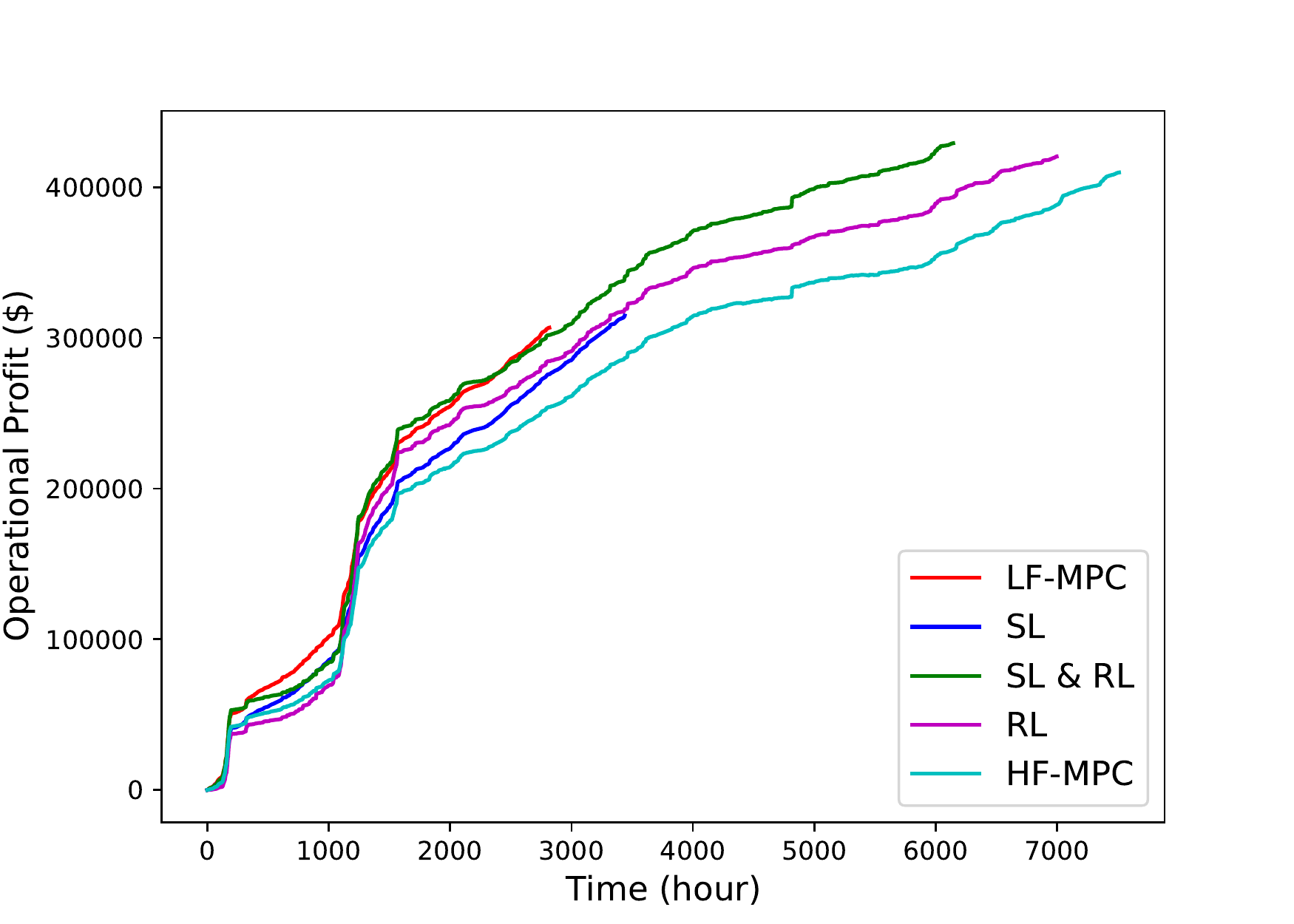}\\
  \caption{Time profiles of operational profit for different FR policies.}
  \label{pic:profit}
\end{figure}

\begin{figure}
  \centering
  \includegraphics[width=0.7\textwidth]{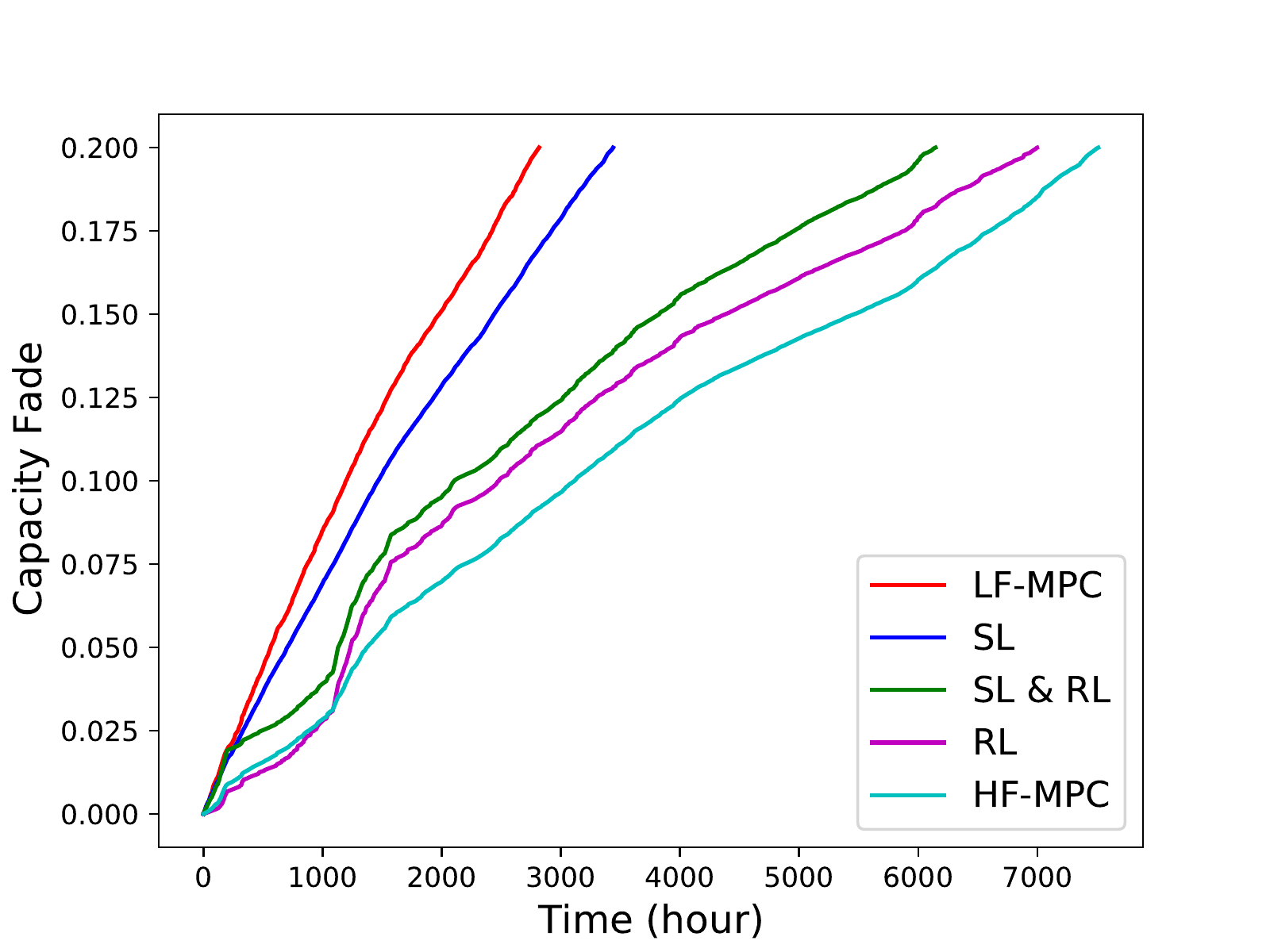}\\
  \caption{Time profiles of capacity fade $C^f$ for different FR policies.}
  \label{pic:capacity_fade}
\end{figure}

Figs. \ref{pic:profit} and \ref{pic:capacity_fade} depict how the operational profit and capacity fade evolve over time under different FR policies. It is clear that the profit of the RL-based policies (SL \& RL, RL) is much higher than that of the LF-MPC and the SL, which justifies the effectiveness of our proposed scheme. Although the HF-MPC can consider the effects of battery degradation and achieve a comparable operational profit, its implementation entails solving a large-scale and highly nonlinear optimization problem online (a highly nonlinear NLP with 34,000 variables). Hence, the online computational cost of the HF-MPC is pretty high. 
In contrast, DNN-approximated policies only need simple function evaluation and are computationally efficient. In addition, for computational tractability, MPC-based schemes can only optimize the cumulative stage rewards over a short prediction horizon. However, RL-based schemes can handle the cumulative reward signals over an infinite horizon. Another advantage of the RL-based policies is that they can consider the effects of battery degradation without having the analytic expressions of the high-fidelity battery model. The proposed RL-based policy is model-free and can be learned from the interaction transition tuples $(x_t,a_t,r_t,x_{t+1})$ as long as a high-fidelity battery simulator is available. Currently, many physics-based electrochemical battery models are available. However, the identification of parameters for those models is a nontrivial task. Within our proposed design framework, the battery model identification is not necessary, and the learning process can be conducted with hardware-in-the-loop simulation.

\begin{figure}
  \centering
  \includegraphics[width=0.75\textwidth]{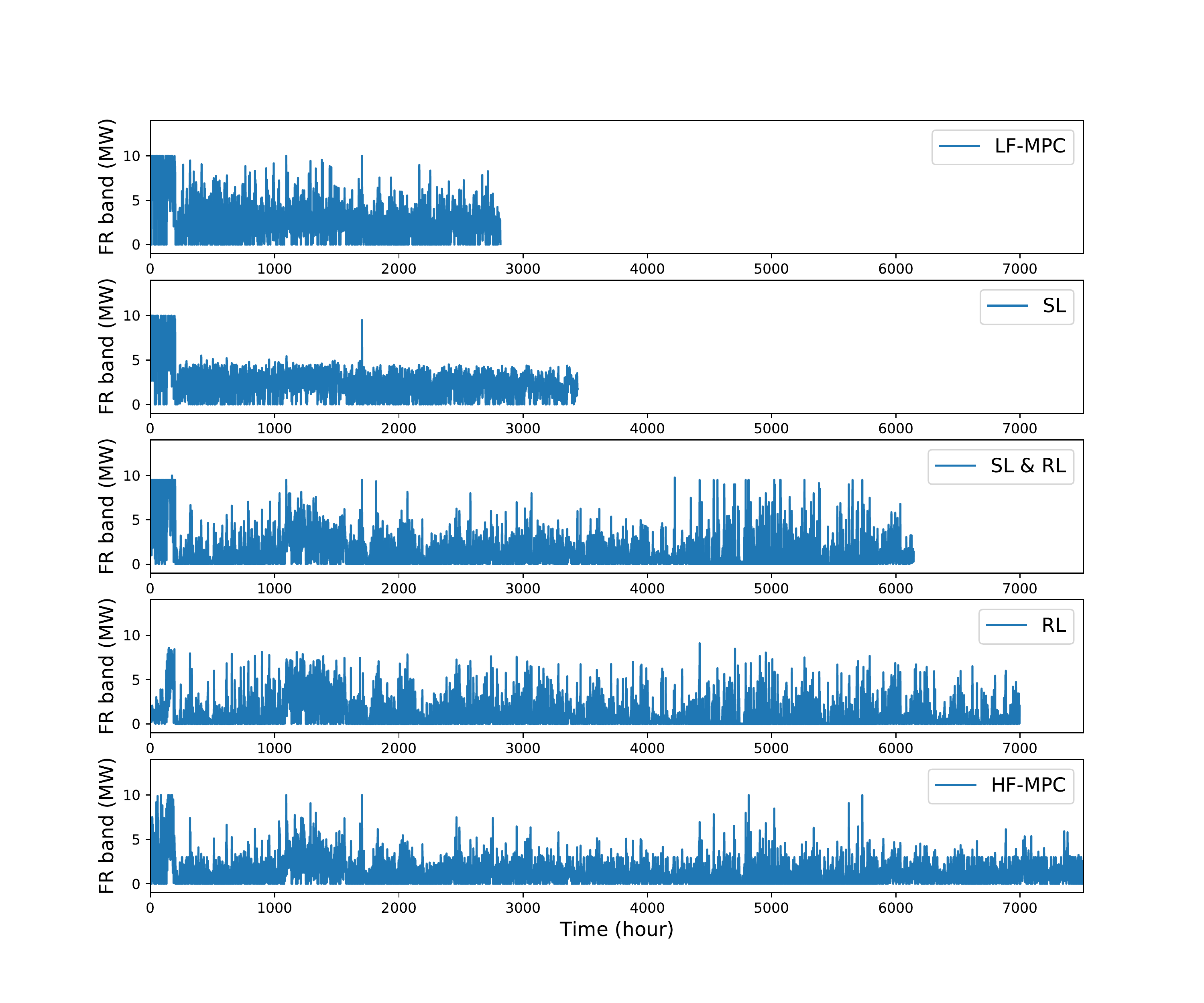}\\
  \caption{Time profiles of FR band for different FR policies.}
  \label{pic:fr_band}
\end{figure}

\begin{figure}
  \centering
  \includegraphics[width=0.75\textwidth]{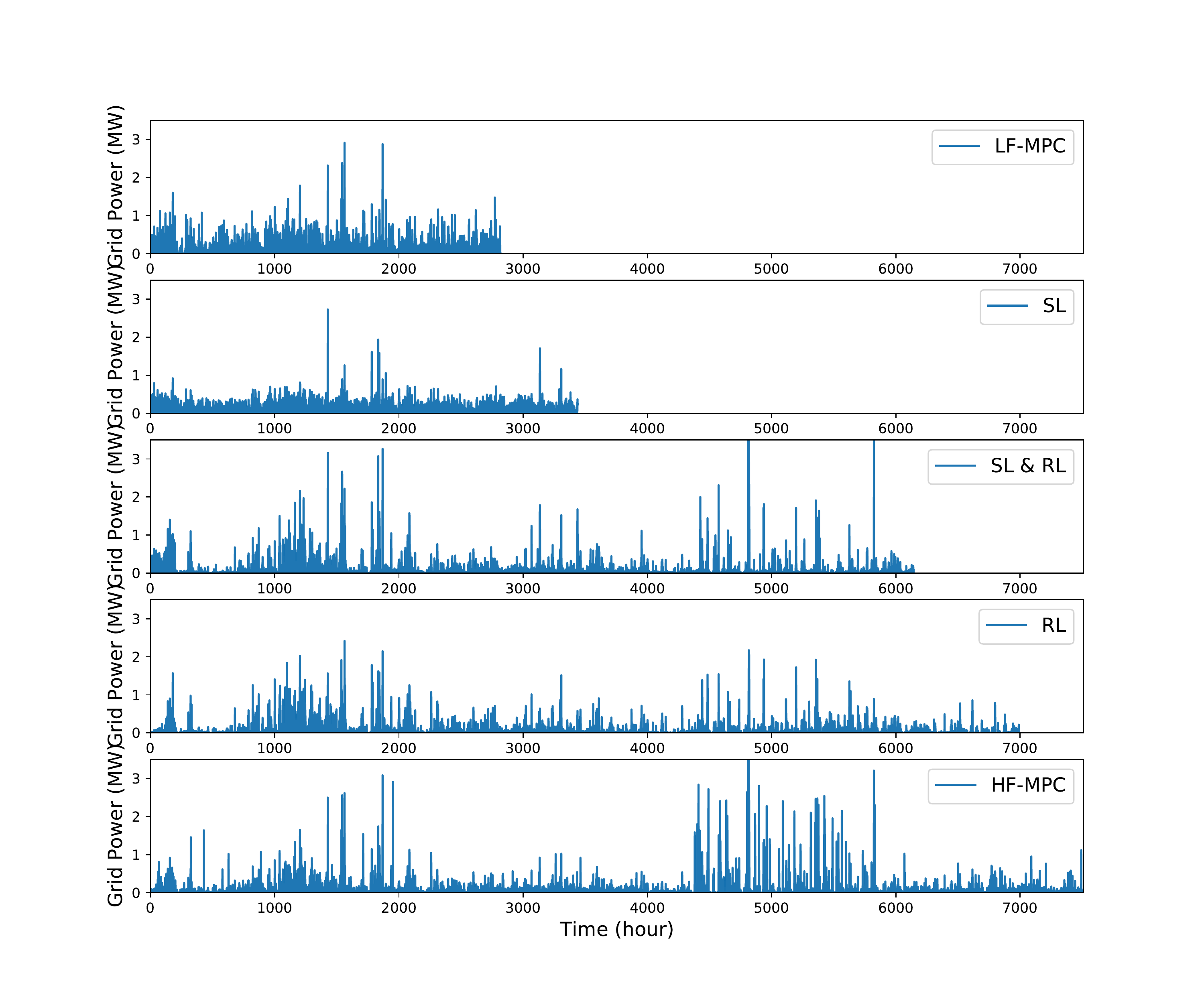}\\
  \caption{Time profiles of purchased power for different FR policies.}
  \label{pic:grid_band}
\end{figure}

It is shown in Fig. \ref{pic:profit} that, by adjusting the DNN-approximated policy $\mu_{\theta}(\cdot)$ via the DPG algorithm, the profit of FR participation is improved remarkably. We attribute this improvement mainly to the extension of the battery's lifetime when using RL-based FR policies since the operational profit of different policies has no significant differences for the first several months and the main difference is in the length of the battery's lifetime. It can be seen from Fig. \ref{pic:capacity_fade} that the capacity fade of RL-based strategies grows much slower than that of the LF-MPC and the SL. For instance, at hour 2818, the LF-MPC has reached its EOL, but the SL \& RL only lost 12\% of its capacity. Figs. \ref{pic:fr_band} and \ref{pic:grid_band} show time profiles of the FR band $F_t$ and the grid band $O_t$ with different FR policies. Compared with the LF-MPC policy, the RL-based policies provide more strategic commitment decisions, and the committed FR capacity is more conservative. This demonstrates that our proposed scheme can effectively avoid aggressive commitment decisions and trade off short-term economic incentives and long-term battery degradation. 

Table \ref{tb:result} summarizes the simulation results of different FR policies, where the cumulative FR band is defined as the total amount of FR capacity committed to the ISO throughout the battery's lifetime. It can be calculated that, compared with the LF-MPC, the SL \& RL policy improves the profit by about 120,000\$, the battery's lifetime by 118\%, and the cumulative FR band by 3\%. This means that, with a comparable cumulative FR band, the RL-based strategy can significantly improve the total profit and extend the battery's lifetime. Furthermore, it can be seen that while the HF-MPC gives the longest battery lifetime among all FR policies, but its profit is lower than that of the RL-based policies. This is because HF-MPC only considers a short prediction horizon, and hence the resulted FR decision is myopic.

In addition, compared with the SL \& RL policy, while the RL policy has a longer battery lifetime, its profit is lower than that of the SL \& RL. This is due to the absence of the SL process when training the RL policy. Adopting the SL process can provide a good initialization for the subsequent RL process. Without a good initialization for the actor $\mu_{\theta}(\cdot)$ and the critic $Q_w(\cdot,\cdot)$, it takes a longer time to learn a policy with high performance (it takes on average 0.93 hour for the SL \& RL to learn for one episode, and 0.95 hour for the SL to learn for one episode). We found that the FR decisions derived via the RL policy in training period can be infeasible quite often and need to be tuned via the backup scheme frequently, which also causes the offline learning process takes longer time. Consequently, the SL process for providing a good initialization for the subsequent RL process is useful and necessary in reducing offline learning time and improving the success rate of learning.


\begin{table}
\centering
\caption{Comparison of different FR policies over one year of operation.}
\label{tb:result}
\begin{tabular}{l|cccccc}\toprule
\hline
 & \makecell{Life time\\(hour)}& \makecell{Revenue\\(\$)} & \makecell{Cost\\(\$)} & \makecell{Profit\\(\$)} & \makecell{Cumulative\\FR band (MW)} & \makecell{Purchased\\power (MWh)}\\\hline
 LF-MPC & 2818 & 350388 & 43624 & 306764 & 7917 & 333\\
 SL & 3439 & 363622 & 48838 & 314784 & 8650 & 360\\
 SL \& RL & 6145 & 492383 & 63244 & 429139 & 8142 & 429\\
 RL & 6997 & 476291 & 55948 & 420343 & 6828 & 390\\
 HF-MPC & 7511 & 474451 & 64728 & 409723 & 9061 & 490 \\
\hline
\end{tabular}
\end{table}

\section{Conclusion}
This paper has presented a deep learning-based battery management strategy for FR leveraging MPC, SL, RL, and high-fidelity battery models. We have shown that, compared with conventional MPC-based approaches, the DNN-approximated FR policy is computationally efficient in online implementation and can also maintain satisfactory performance. In addition, by taking advantage of RL algorithms, the proposed scheme is able to trade off short-term economic benefits and long-term battery degradation to prolong the battery's lifetime and improve the total profit of FR participation. Furthermore, the proposed learning-based design does not require analytic expressions of the high-fidelity battery model and can adapt the DNN-approximated FR policy by learning from the data of interacting with the high-fidelity battery simulator. Future works include considering more complicated energy management settings and designing data-efficient learning schemes.

\begin{acknowledgement}
Yankai Cao acknowledges financial support by the Natural Sciences and Engineering Research Council of Canada under grant RGPIN-2019-05499. We gratefully acknowledge the computing resources provided by Compute Canada (www.computecanada.ca).
\end{acknowledgement}

\bibliography{achemso-demo}

\end{document}